**Importance of method validation: Implications of non-correlated observables in sweet taste receptor studies**

Rani P Venkitakrishnan[1] and Manojendu Choudhury[2]

**Affiliation**:

[1]Dept. of Chemistry, Center for Basic Excellence in Science (UM-DAE CBS), Univ. of Mumbai, Mumbai, Kalina – 400098

[2]Dept. of Physics, Center for Basic Excellence in Science (UM-DAE CBS), Univ. of Mumbai, Mumbai, Kalina – 400098

Corresponding author: rani.parvathy@cbs.ac.in, phone +91-9967803377, fax: +91-22-26524982

**Running Title**: Non-correlation of observables in calcium flux assay of sweet taste receptors

**Abstract**

Validation of research methodology is critical in research design. Correlation between experimental observables must be established before undertaking extensive experiments or propose mechanisms. This article shows that, observables in the popular calcium flux strength assay used in the characterization of sweetener-sweet taste receptor (STR) interaction are uncorrelated. In pursuit to find potential sweeteners and enhancers, calcium flux generated via G-protein coupling for wildtype and mutant receptors expressed on cell surface is measured to identify and localize sweetener binding sites. Results are channeled for sweetener development with direct impact on public health.

We show that flux strength is independent of EC50 and sweet potency. Sweet potency-EC50 relation is non-linear and anti-correlated. Single point mutants affecting receptor efficiency, without significant shift in EC50 have been published, indicating flux strength is independent of ligand binding. G-protein coupling step is likely observed in the assay. Thus, years have been spent generating uncorrelated data. Data from uncorrelated observables does not give meaningful results. Still, majority of research in the field, uses change in calcium flux strength to study the receptor. Methodology away from flux strength monitor is required for sweetener development, reestablish binding localization of sweeteners established by flux strength method. This article serves to remind researchers to validate methodology before plunging into long term projects. Ignoring validation test on methodology, have been a costly mistake in the field. Concepts discussed here is applicable, whenever observable in biological systems are many steps moved from the event of interest.

# Introduction

The sweet taste receptor, binds to small molecule sweeteners and large molecular weight sweet tasting proteins, and elicits sweet taste. It is an important research target, in sweetener development, synergy studies to reduce total amount of sweetener in foods and for health benefits to reduce obesity and diabetes. *Tas* I genes encode protein sequences for T1R2-T1R3 sweet taste receptor and T1R1-T1R3 umami taste receptor [1-5]. Umami taste receptor is responsible for taste of MSG. Many present working hypotheses in the field is derived from concepts in mGluR [6-10], the best characterized receptor of Class C GPCR. Publications in the field popularly use G-protein generated cell based activity assay to monitor receptor-ligand binding. Experiments involve co-transfection of pairs of wildtype, mutant or chimeric GPCR cDNAs into heterologous cell lines. Activity of the receptor expressed in eukaryotic cells, is monitored by $Ca^{2+}$ influx for taste receptors or $IP_3$ flux for mGluRs. Receptor couple to G-proteins, which effect PLC/PKC levels, causing calcium flux across cell membrane. Data for wild type and mutants are compared and translated to ligand binding or mechanisms [11-21]. Increase in calcium flux strength is considered an indicator of enhancer and decreased flux an indicator of loss of ligand-receptor interaction to map binding sites.

The system has three observables (i) EC50 (ii) flux strength and (iii) sweet potency. If researchers base their conclusions on change in flux strength, a correlation must exist between the flux strength and the other variables for meaningful results. If empirical correlations between observables are assumed before start of experiments, generated data must reinforce the assumptions. In studies involving complex systems, it is acceptable at first to derive empirical correlations that are statistically valid, for guidance to obtain scientific system description. Empirical correlations, by themselves are strengthened and redefined when experimental data becomes available. Then link between empirical and scientific correlation is explained as a mechanism or pathway. If empirical assumptions are not valid, an alternate methodology is required.

**Experimental Procedure**

In cell calcium flux assay has three variables, EC50, flux strength and sweet potency. Data published by independent groups in the field was tabulated and analyzed by two methods. First, an XY plot between the variables was generated. Then, three parameter Spearman Rank Correlation (SRC) test, was performed on the experimentally generated and established data points, to determine correlation between the three variables and the strength of association. An in-house program written in Fortran 77 for data analysis adapting Macklin's method prescribed in 1982 [22] was used for correlation tests. The program established to make prior mathematical calculations for publications was used to generate the correlation coefficients [23, 24]. This brings out interdependencies among variables when more than two variables are present.

**Results**

Table 1 summarizes reported values of sweet potency, EC50 and flux strength for several sweeteners from multiple independent publications. Sweet potency is determined via taste panels. EC50 is the apparent dissociation constant ($K_{d-apparent}$). $K_{d-apparent}$ differs from receptor-ligand dissociation constant and does not represent the $K_d$, derived from direct observables. Plot of observed EC50 against sweet potency, shows inverse non-linearity, from Figure 1. Low potency-natural sweeteners and high potency-artificial sweeteners are separated by a sharp bend. Most commonly used sweeteners fall under the bend region of the curve. Collated flux strength data, indicate random change and show no visual correlation to EC50 or sweet potency as seen from Figure 2. The most potent sweetener in market, neotame, does not show increased flux compared to other sweeteners in the published studies, raising concerns on flux strength as a valid variable.

When a clear correlation is not established by an initial inspection of the data or when there is no mathematical functions which fit the data, as in Figure 1, independent mathematical correlation tests are essential to establish validity. For correlation among three variables, the null hypothesis is that the correlation between two variables arises entirely from those of the third variable with the first and second variable separately. Modulus of the SRC lies between 0 and 1, the negative value signifies anti-correlation. Significance level associated with correlation between, two variables, independent of third variable is given by D-parameter. If the null hypothesis that the correlation between the first two variables arises entirely from the correlation of the, third variable, with the first and second independently is true, then, the D-parameter, is normally distributed about zero with unit variance. Results of the correlation test are given in Table 2. From the table, we see strong anti-correlation is between EC50 and sweet potency with correlation coefficients of -0.95 and -0.98. For both data sets from Table 1, EC50 is strongly anti-correlated to sweet potency, and not calcium flux strength. Since the strong anti-correlation is non-linear, as seen from Figure 1, extreme caution is to be exercised when mutation studies are interpreted for sweetener development.

Research has yielded single point mutants exhibiting decreased receptor activity, without significant change in EC50 for many sweeteners [13, 21]. Wildtype and mutant receptors exhibit similar activity profiles, with drastic change in maximal response (Supplemental Figure 2 of reference 13). This means receptor bind ligand with similar affinity, yet, is not activated to it's complete potential. From Figure 1, we know, change in sweetness manifests as non-linear shift in EC50. Calcium flux is a measure of efficacy (efficiency) of the receptor. Coupling of the receptor to the G-protein, which is the subsequent step after ligand binding is observed and the most possible rate-limiting step in the process.

**Discussion**

The above results establish, flux strength is independent of EC50 and also not a monitor of ligand binding, unlike previously assumed. To make meaningful conclusions, based on change in flux strength, a linear relationship with flux strength or EC50 is a must. For successful sweetener or enhancer development, correlation between variables (observables) is required. For sweet taste receptor, this is clearly not achievable by monitoring receptor activity due to non-existing correlation between the observables. Even if G-protein cascade starts due to receptor-ligand interaction, change in flux strength is not governed by ligand-receptor interaction. Rate determining step in a cascade process is the slowest step in the cascade. It also governs calcium flux changes and an unknown in cell based flux assay. Such complex receptor systems have (i) ligand interaction & affinity which are internal properties and (ii) external parameter of efficacy (efficiency), which is the efficiency of coupling to downstream events that are triggered. Single point mutant with decreased activity without significant change in EC50 have been reported for umami taste receptor [25] too, the other taste receptor member of class C GPCR. This shows, for both sweet and umami receptors, downstream flux assay is a monitor of G-protein coupling instead of ligand binding. The G-protein cascade can be summarized as in the following equation, with each reaction characterized by individual rate constants and equilibrium constants. The third step involves $IP_3$ binding to the $IP_3$ receptor and causing calcium influx, which is the change that is measured. The cumulative kinetics of the cell cascade is observed in cell based assays.

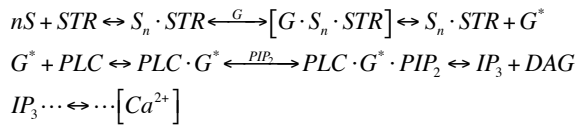

$$nS + STR \leftrightarrow S_n \cdot STR \xleftarrow{G} [G \cdot S_n \cdot STR] \leftrightarrow S_n \cdot STR + G^*$$
$$G^* + PLC \leftrightarrow PLC \cdot G^* \xleftarrow{PIP_2} PLC \cdot G^* \cdot PIP_2 \leftrightarrow IP_3 + DAG$$
$$IP_3 \cdots \leftrightarrow \cdots [Ca^{2+}]$$

Current working concept suggests, increase in secondary structure content for the extracellular domain upon ligand binding, similar to mGluR. However, circular dichroism studies by independent groups show, decrease in secondary structure content on ligand binding [26, 27], opposite of the present working concept. Mathematical equations to describe multi-substrate systems can be applied if the observable is the direct consequence of ligand interaction [28-30]. It is not recommended for $Ca^{2+}$ or $IP_3$ flux assay analysis, since observable is many steps removed from the initial ligand binding step. Cooperativity analysis, which generate Hill coefficient (stotiometry ratio of receptor to ligand) is based on direct observable measurements.

Goudet et al [31] have observed, diminished efficiency for mutant mGluR lacking the extra cellular domains with inhibitors, suggesting contribution from extracellular and transmembrane domains, towards receptor activity. Likely, the same holds for sweet taste receptor due to the following. Since most artificial sweeteners elicit bitter tastes by interaction with bitter receptor, which lack the extra cellular domains, the sweetener candidates likely bind to the transmembrane region of bitter taste receptors. Natural sweeteners with low activity do not elicit other tastes are seen in the low potency region. High potency artificial sweeteners, capable eliciting alternating tastes are clustered in the bend region onwards as seen from Figure 1. It is possible that, ECD and transmembrane regions have varied contribution towards receptor activity in sweet taste receptor too, with distinct contributions from the two activating regions in the receptor, with the low potency and high potency regions separated by the sharp bend in Figure 1. The sharp bend does not conform to any continuous mathematical function, indicating more than one state in the receptor activation. This is in accordance to the fact that, if a mathematical function can fit the complete set of data then correlation becomes superfluous. More than one function operating at different ranges are required to fit such data curves, in accordance with multiple possible states of receptor activation. Here too it is important that mathematical and physical models go hand in hand.

IP$_3$ or calcium fluctuations are the overall non-synchronous out of phase response of an ensemble of cells to the stimuli and mediated by many cellular processes. This is a property of the cell type used in the studies and not a monitor of molecular or conformational mechanism. Attempts to find selective mutations for the sweet panel produce complex effects on most sweeteners [16, 17, 19, 32]. Mutations in domain-interfaces, hinge, or linker regions effect receptor activity.

Observables at the level of action potential, such as firing rate and inter-burst rates, are known to differentiate between sweeteners [33]. They are independent of stimulus intensity, and possible candidates for alternate measurements. Advances in technology for detection of cell activities, has led to a tremendous increase in use and popularity of cell based assays to probe protein processes. Concepts discussed above, are applicable, whenever, measurement is made on an observable, steps removed from the event of study.

Scarcity of activity curves in publications and activity representation as bar chart is a drawback in the field. This obscures important details to the reader. Take home message is, establishing correlation between observables are crucial in biological assays to establish biophysical parameters of receptors. What more can be done to inform researchers and rectify the issue? Why did many overlook to establish correlations, role of rate limiting step was excluded is a lingering question. Why did peer-review break down and not detect this for so long? More than required time, money and brainpower have been expended in generating uncorrelated experimental results.

**Acknowledgement**

We thank the generous and kind Late Prof. W. W. Cleland, with whom the analysis was conducted. We thank Prof. G Krishnamoorthy and Prof. TFJ Martin for discussions.

**Author Contribution**

Manojendu performed correlation tests, analysis and significance of test. Rani contributed to data analysis and write up, late Prof. Cleland geared analysis and guidance.

**Footnotes**

Key words: Sweet taste receptor, T1R2, T1R3, calcium flux, G-protein, cell activation, GPCR, sweetener, enhancer, cooperativity, correlation test, Spearman rank correlation, metabotropic glutamate receptor (mGluR)

**Figure Legends**

**TABLE 1**

Sweet potency, EC50 and calcium flux strength of various sweeteners, from published data are tabulated [13-15, 26, 34, 35]. Flux strength from different data sets cannot be quantitatively compared. Flux strength is not linear with respect to EC50 or sweet potency as observed from experimental results ([1]Koizumi 2011, [2]Zhang 2010). T1R2-R217A does not change EC50 values significantly, but decreases activity of the receptor for several sweeteners. Flux strength for wildtype is in parentheses. The R217A mutant also shows non-linear flux strength response.

**TABLE 2**

(A) Results of correlation test from column 4 data of Table 1. Correlation between EC50 and flux strength does not exist as seen from the correlation coefficient. Potency: EC50 is strongly anti-correlated with negative correlation coefficient of -0.98, with a chance probability of $10^{-8}$ ~ 0 (probability of the correlation occurring by chance), and D-parameter is between 1 and -1, therefore the anti-correlation doesn't depend on the third parameter Z.

(B) Results of correlation test from column 5 data of Table 1. The correlation between EC50: flux strength and flux strength: potency is very weak, and can be attributed to the third parameter, but the significance of these correlations is very low. The weak correlation observed between EC50:flux strength and potency: flux strength likely arises due to the strong anti-correlation between EC50:potency at -0.98, as per the D-parameter values in the table.

**FIGURE 1**

Plot of EC50 in mM against sweet potency. EC50 the apparent $K_d$ of activation, does not have a linear correlation with sweet potency. The hyperbolic-like profile with sharp bend, consist of two regions. Inset is the zoomed in region of the bend region. (a) neotame, (b) Monellin, (c) Thaumatin, (d) NHDC (e) Brazzein, (f) sucralose, (g) saccharin, (h) stevioside, (i) acesulfame-k, (j) aspartame, (k) cyclamate, and (l) sucrose. Right side is populated with low potency sweeteners and left side with high potency sweeteners. Many artificial sweeteners elicit bitter tastes. Since bitter taste receptors lack the extra cellular domains, they are activated via transmembrane region. Natural sweeteners with low activity do not elicit other tastes.

**FIGURE 2**

Calcium flux strength varies randomly with respect to EC50 and sweet potency, without correlation. a & b are derived and plotted from [36], c & d from [25]. Flux strength varies randomly over EC50 and sweet potency. Sweeteners and potencies in a & c are 1-sucrose, 30-cyclamate, 200-acesulfame K & aspartame, 250-stevioside, 300-saccharin, 600-sucralose, 1500-Brazzein, 2000-thaumatin, 8000-neotame. Compounds & EC50 (µM) in b & d are, 0.2-D-trp, 0.83-neotame, 1.2-Thaumatin, 80-sucralose, 190-saccahin, 25-0stevioside, 750-aspartame,1900-cyclamate.

**Table 1**

| Ligand | Sweet Potency | EC50 (µM) | Activity[1] | Activity[2] | R217A EC50 | R217A Flux |
|---|---|---|---|---|---|---|
| sucrose | 1 | 10000 | 16.5 | 21 | | |
| cyclamate | 30 | 1900 | 19 | 9 | 1120 | 55 (110) |
| aspartame | 200 | 750 | 24 | 20 | | |
| acesulfame K | 200 | 540 | 13.5 | | | |
| stevioside | 250 | 250 | 11 | 11 | | |
| saccharin | 300 | 190 | 19 | 14 | | |
| sucralose | 600 | 80 | 20 | 10 | 120 | 60 (103) |
| Brazzein | 800 | 16 | 57 | | 22 | 43 (118) |
| NHDC | 1500 | 0.2 | 46 | 17 | | |
| Thaumatin | 2000 | 1.2 | 57.5 | 24 | | |
| Monellin | 3000 | 11 | | | 14 | |
| neotame | 8000 | 0.83 | 18 | 15 | 0.67 | 45(100) |
| D-trp | | 2.09 | 24 | | | |

Table 2A

| observable | SRC | chance probability | D-parameter |
| --- | --- | --- | --- |
| potency : EC50 | -0.984018147 | 4.71811887E-08 | -6.05579281 |
| EC50 : flux strength | -0.513698637 | 0.106024571 | -0.689184785 |
| flux strength : potency | 0.481641352 | 0.133597195 | -0.416386008 |

Table 2B

| observable | SRC | chance probability | D-parameter |
| --- | --- | --- | --- |
| potency : EC50 | -0.950000048 | 8.76249833E-05 | -4.08812809 |
| EC50 : flux strength | -0.116666675 | 0.765008211 | 0.187551886 |
| flux strength : potency | 0.149999976 | 0.700094342 | 0.283922881 |

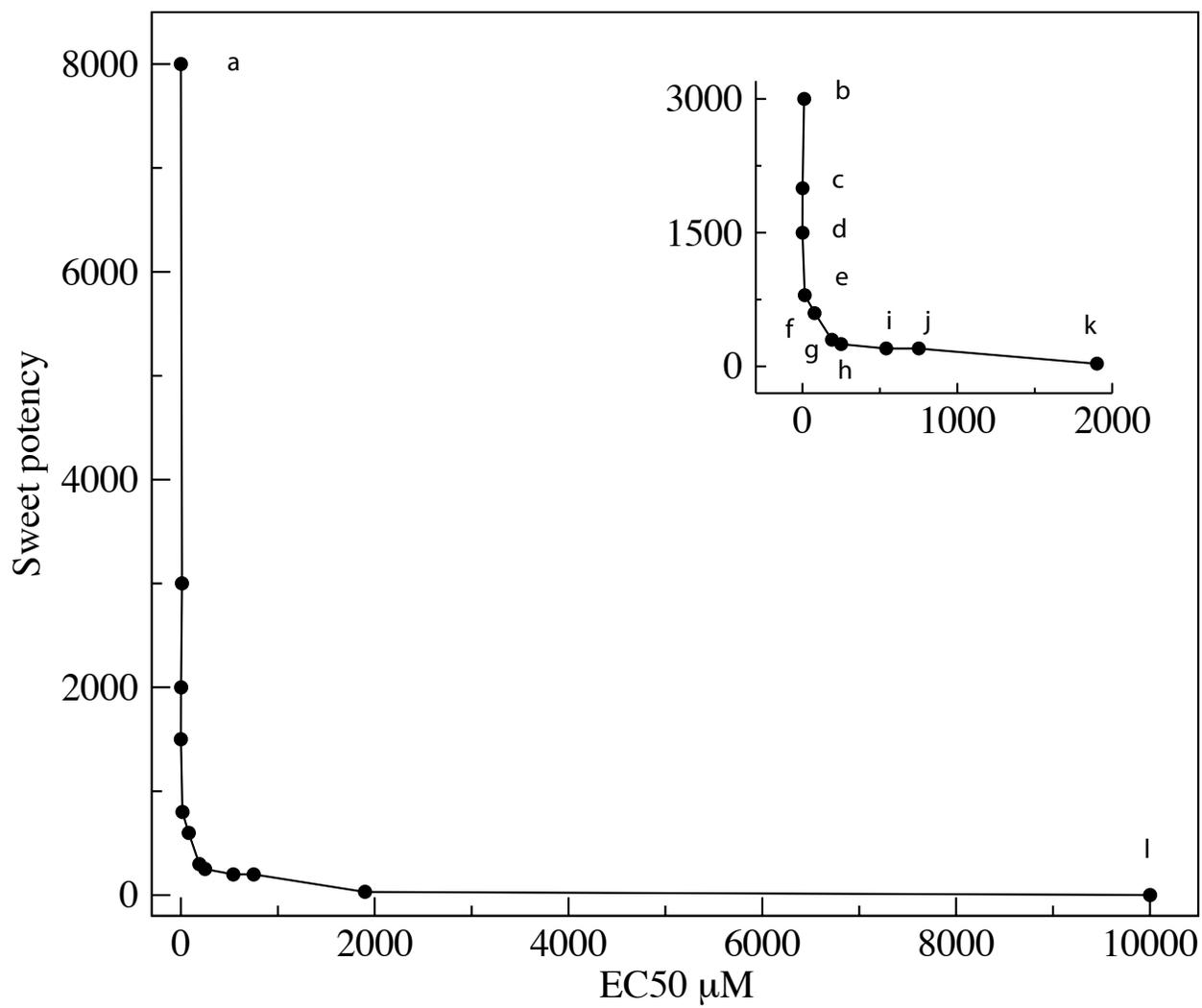

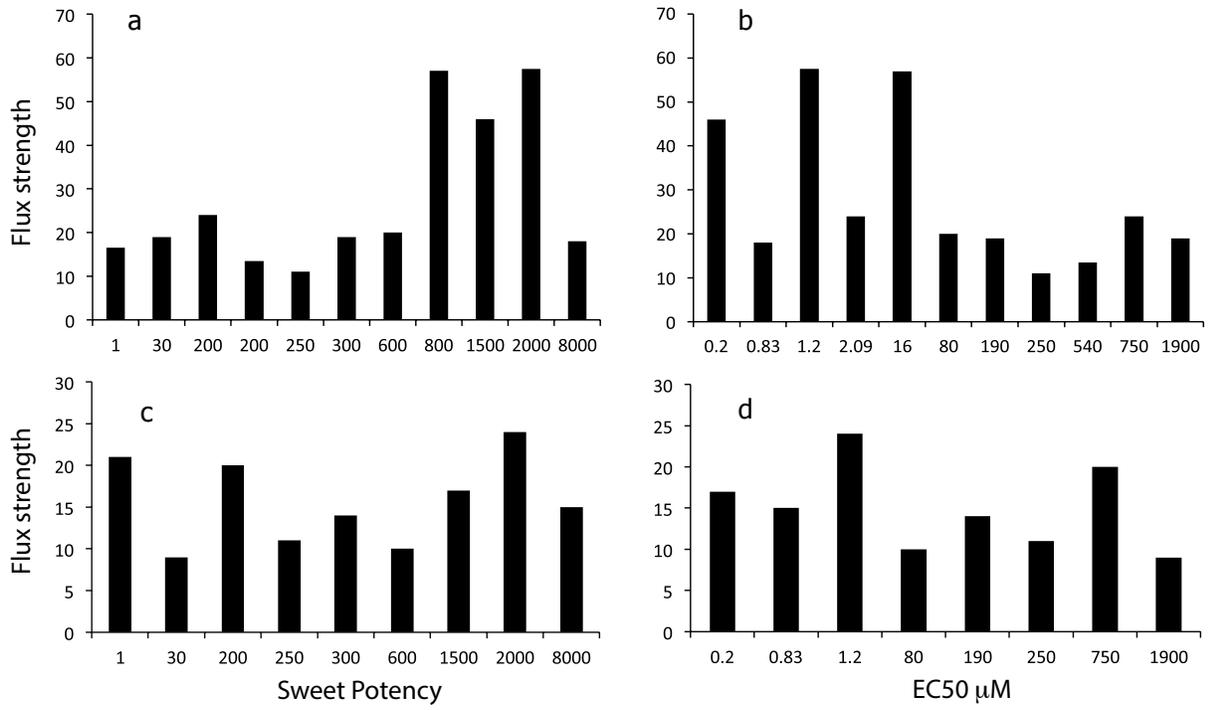

**Figure 2**